\documentclass[journal]{IEEEtran}
\usepackage{mathptmx}
\usepackage{helvet}
\usepackage{courier}
\usepackage{type1cm}
\usepackage{makeidx}
\usepackage{graphicx}
\usepackage{amsmath}
\usepackage{multicol}
\usepackage[bottom]{footmisc}
\usepackage{subfigure}
\usepackage{multicol}
\usepackage{rotating}
\usepackage{array}
\usepackage{algorithm,algorithmic}
\usepackage{amsmath}
\usepackage{amsfonts}
\usepackage{stmaryrd}
\usepackage{bm}
\usepackage{color}
\usepackage[nospace,noadjust]{cite}

\graphicspath{{figures/}}

\ifCLASSINFOpdf

\else

\fi

\begin{document}

\title{Effective Data Aggregation Scheme for \\Large-scale Wireless Sensor Networks}

\author{M. Mehdi Afsar 
\thanks{M. Mehdi Afsar is with the Young Researchers and Elite Club, Mashhad Branch, Islamic Azad University, Mashhad, Iran. (Email: m.afsar@qiau.ac.ir)}
}



\maketitle

\begin{abstract}
 Energy preservation is one of the most important challenges in wireless sensor networks. In most applications, sensor networks consist of hundreds or thousands nodes that are dispersed in a wide field. Hierarchical architectures and data aggregation methods are increasingly gaining more popularity in such large-scale networks.  In this paper, we propose a novel adaptive Energy-Efficient Multi-layered Architecture (EEMA) protocol for large-scale sensor networks, wherein both hierarchical architecture and data aggregation are efficiently utilized. EEMA divides the network into some layers as well as each layer into some clusters,  where the data are gathered in the first layer and are recursively aggregated in upper layers to reach the base station.  Many criteria are wisely employed to elect head nodes, including the residual energy, centrality, and proximity to bottom-layer heads.  The routing delay is mathematically analyzed.  Performance evaluation is performed via simulations which confirms the effectiveness of the proposed EEMA protocol in terms of the network lifetime and reduced routing delay. 
\end{abstract}


\IEEEpeerreviewmaketitle

\section{Introduction}
\label{sec:intro}

Recent advances in miniaturization and wireless communications have enabled making the micro sensors with limited processing and communicating capabilities.  Large-scale WSNs consist of a large number of sensor nodes, thousands or millions~\cite{Akyildiz2002}, scattered in a wide field and provide different types of applications~\cite{Yick2008}. WSNs are drastically energy-constrained  so that energy preservation is one of the most important challenges in these networks.
Accordingly, the long lifetime is usually considered as a desired goal in the design level of such networks. On the other hand, the large number of the nodes imposes some overheads, including increased routing table size and delay; and makes the scalability issue difficult in such large networks. 

As shown in~\cite{Kleinrock1977}, hierarchical architectures are effective approaches in making large traditional networks scalable by reducing the  size of the routing tables.  Although WSNs seem different from traditional networks (even Mobile Ad hoc NETworks (MANETs)), hierarchical architectures shown to well match these large networks. At the same time, since WSNs are data-centric~\cite{survey-karaki} and the notable value of the data produced by sensors are the same, data aggregation is an effective approach to reduce the load, and as a result, helps the network to be more energy-efficient.  

In the last decade, although clustering the nodes has been extensively investigated for WSNs, hierarchical multi-layered architecture has not been properly explored.  Hence, in this paper, we analyze the impact of an adaptive hierarchical multi-layered architecture on the energy-efficiency issue of large-scale WSNs. The proposed Energy-Efficient Multi-layered Architecture, called EEMA, divides the entire network into some layers, as well as each layer into some clusters. The head of each cluster is selected based upon a hybrid of residual energy, centrality, and the location of node. The data are gathered in the first layer, and  are hierarchically aggregated in the next layers to reach the base station (BS).  Adaptive clustering is employed to achieve load balancing among all the nodes, and consequently, improve the lifetime.   We also analyze the routing delay mathematically and show that EEMA significantly improves which.  Furthermore, we mathematically show that the elected head nodes in the extra layers could participate in data acquisition so that the coverage is conserved as good as clustering approaches.  The performance of EEMA is validated through simulation.
 
The rest of this paper is organized as follows. Related work is discussed in section~\ref{sec:re-wo}.  Preliminaries about the used system in this paper are discussed in section~\ref{sec:net-mod}.  Section~\ref{sec:EEMA} explains the proposed EEMA protocol in detail. Performance evaluation and experiments are presented in section~\ref{sec:exp} and the paper is concluded in section~\ref{sec:con}.

\section{Related Work}
\label{sec:re-wo}
The following presents the most popular and recent clustering approaches.  As early attempts in the area of clustering in WSNs, LEACH~\cite{Heinzelman2002} has been proposed by Heinzelman {\it et al.} which is an application-specific protocol and uses a random probabilistic approach for CH election.  Another baseline clustering approach is HEED~\cite{Younis2004} that is iterative-based and uses a hybrid of the node residual energy and communication cost (such as AMRP or node degree) to select the CHs. More precisely, the residual energy is used as primary parameter to select an initial set of the CHs ({\it tentative CHs}), and then AMRP (minimum power level required by a node to communicate with its CH) or node degree is used as secondary parameter to break ties.  DWEHC~\cite{Ding2005} is an improvement on HEED which utilizes a weight-based approach to form clusters.  This weight is a function of the residual energy of the  candidate node, and also, whose proximity to its neighbors.  Furthermore, DWEHC supports multi-hop intra-cluster communications in order to achieve a better energy-efficiency.
 We will compare our work with these baseline clustering approaches in section~\ref{sec:exp}.  ACE~\cite{ACE} is based on emergent algorithm and factors the node degree to form clusters with reduced overlapping.  FLOC~\cite{FLOC} uses the state transitions in order to select the CHs.   Overlapping multihop clustering (KOCA) is proposed in~\cite{koca2009} which generates connected overlapping clusters that cover the entire sensor network with a specific average overlapping degree.  
In KOCA the load is distributed uniformly among all the equal-size clusters. 
In~\cite{soro2005}, an unequal clustering size (UCS) scheme is proposed to prolong the network lifetime.  the UCS protocol utilizes a hierarchical clustering approach which provides a two layer clustering, based on the distance of CHs to the BS. In other words, the closer clusters to the BS are smaller in size than other farther ones.  
Similar approach is used in~\cite{UCR} in which based on the distance to the BS, an unequal clustering called EEUC is proposed.
An energy-efficient clustering (EC) solution is proposed in~\cite{EC2011} in which the clusters size is related to the hop distance of the nodes to the BS.  Similar to unequal clustering approaches, the closer clusters to the sink are smaller in size, and the network lifetime of all the clusters is balanced.   
In~\cite{load2013}, a load-balanced clustering algorithm on the basis of their distance and density distribution has been proposed.  A ZigBee-like addressing scheme is proposed in~\cite{pcc2012} where using a distributed formation, the paths are automatically separated from the clusters.  The main advantage of this method is the low generated overhead in address-based routing.
LCM~\cite{lcm2013} is a link-aware clustering approach for WSNs in which the CHs  are elected by evaluating the status of the nodes and the conditions of links. 
In~\cite{Wei2011}, a centralized clustering approach based on neighbors (EECABN) is proposed in which  the CHs are elected based on a weight that includes several factors, like the residual energy and distance of the node to its neighbors  as well as the BS.  And finally, EEDC~\cite{EEDC} uses two criteria for CH election: local competition and distance condition. In the local competition criterion, the nodes compete with one another in a predefined range ($R_{comp}$)  to select the nodes with the highest residual energy as the CH candidates (CCH). When a proper set of  nodes is selected as the CCHs, the algorithm checks if the selected CCHs have enough distance to one another, so the CHs are distributed evenly across the network. The CCHs with a greater (or equal) distance than a threshold distance, $D_{thr}$, are selected as new CHs.  We also compare our approach with EEDC.

In addition to clustering, some popular hierarchical multi-layered architectures has been proposed in some research~\cite{hierarchical2001,EEHC2003,karaki2004,karaki2009}. Hierarchical power-aware routing was proposed in~\cite{hierarchical2001} which divides the network into some groups of nodes as zones. In the routing process, each zone decides to route a message hierarchically along other zones, in order to maximize the network lifetime. The algorithm considers a trade-off between minimizing the energy consumption in the entire network and maximizing the minimal residual energy of the network. 
Another good illustration of hierarchical clustering with main focus on the longevity of the network is EEHC~\cite{EEHC2003}. EEHC selects the CHs by a probability proportional to the density of neighboring nodes within the desired range of the node. Generally, operations in EEHC are classified into initial and extended stages. In the initial stage, the data are gathered and aggregated by the CHs; afterwards, in the extended stage, the data are aggregated and transmitted to the BS among the CHs through a hierarchical multi-tiered path.  EEHC indicates that the hierarchical multi-layered architecture effectively improves the energy efficiency of large-scale WSNs.  However, EEHC does not consider the energy reserve of the nodes.
In~\cite{karaki2004,karaki2009} a hierarchical scheme is proposed to define the most appropriate aggregation points in the network.  LAs (Local Aggregator) is used in the work to aggregate the sensed data of the regular nodes, as well as MAs (Master Aggregator) to aggregate the data of several LAs.  Using heuristic algorithms, the work tries to find the minimum number of the data aggregation points in order to maximize the network lifetime.  However, the work has no indication that the aggregators are elected based on what approach. More importantly, using heuristic approaches has practically problem, where distributed approaches are needed for WSNs.

Furthermore, hierarchical multi-parent data aggregation framework is proposed in~\cite{alemu2010} which uses the CH election algorithm of LEACH in electing two heads for each cluster.  The work focuses on recovering the errors of the nodes, where all operations of the approach are the same as LEACH so that it conserves the problems of  LEACH (i.e. random CH election).  Another hierarchical cluster-based scheme is HRDD~\cite{HRDD} that uses a large-scale WSN with multiple mobile BS.  First clusters are formed using Max-Min D-cluster algorithm and then the data are sent to the mobile BS that the request is received from.  However, the approach has some problems: (1) the approach is so complicated; (2) used clustering approach is borrowed from ad hoc networks which is not suitable for WSNs; (3) data aggregation is not used efficiently.
And finally, another hierarchical two-tier approach is~\cite{bulgaria2012} that aims at balancing energy consumption among all the nodes.  The main idea in which is to consider the residual energy of the current node and needed energy to communicate with next-hop node.  The work is a simple modification over existing popular approaches~\cite{Heinzelman2002,SEP2004}.  In contrast, EEMA solves all of these problems utilizing adaptive clustering, effective data aggregation, and energy-aware head election.

\section{Preliminaries}
\label{sec:net-mod}
In this section, we clarify our used network model. We adopt the following assumptions about the used network:
\begin{itemize}

\item	$N$ nodes are randomly and uniformly dispersed in a square field of size $M \times M$.
\item	All the nodes and the BS are stationary. 
\item	All the nodes can use power control for different distances from the transmitter to the receiver.
\item	All the nodes are location unaware (i.e. they are not equipped with the GPS-devices). 
\item	All the nodes are homogeneous (all capacities).

\end{itemize}

The above assumptions are reasonable and usual in many applications, and also which makes our simulation easier and more real. Note that EEMA does not require a fully-synchronized sensor network and local synchronization might be achieved by exchanging a few packets~\cite{zhu}.

\begin{figure*}[t]
\centering
	\subfigure[The flat architecture without data aggregation.]{
\includegraphics[width=0.45\linewidth]{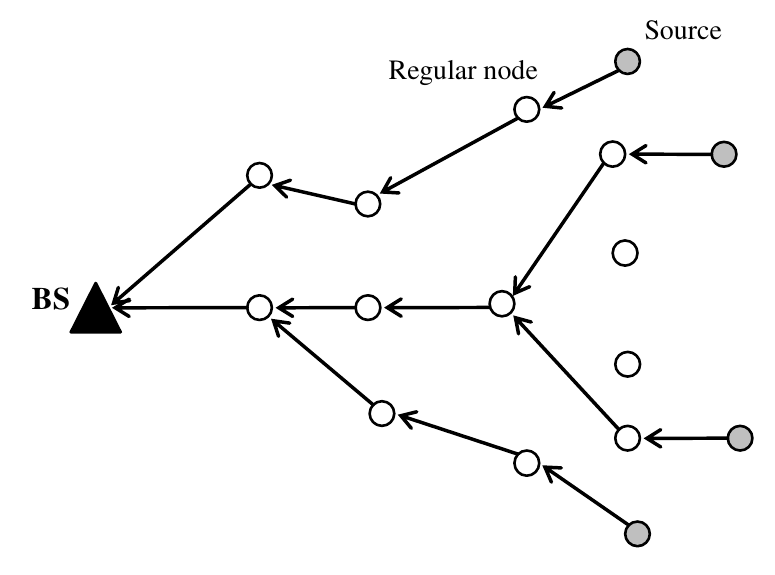}
	    \label{fig:WDA}
	}
	\subfigure[The three-layered EEMA architecture with effective data aggregation.]{
\includegraphics[width=0.45\linewidth]{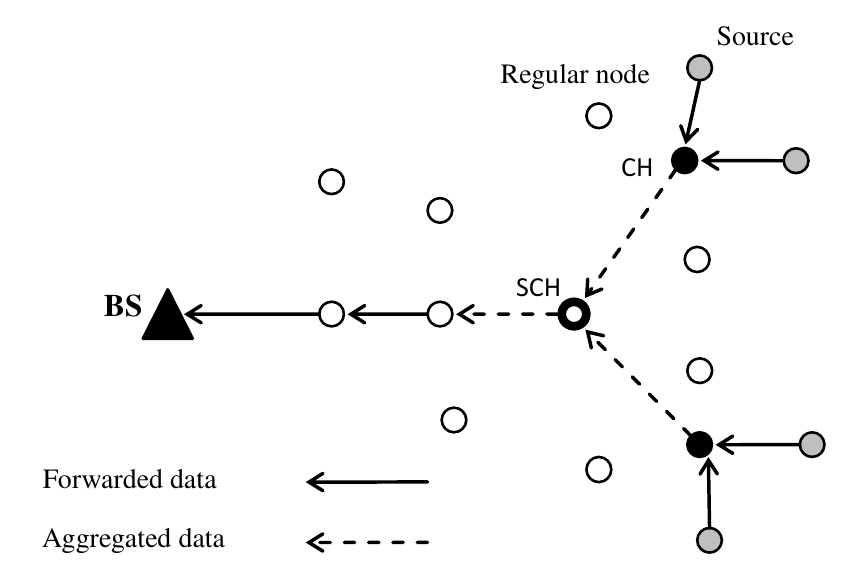}
	    \label{fig:Da}
	}
	\caption{A simple case study; a WSN with/without data aggregation.}
    \label{fig-DA}
\end{figure*}

As discussed in~\cite{impact}, data aggregation is effective when the network size and the number of sources (here source is a node that has some data to send) are large, sources are relatively close to each other, and their distance to the BS is far.  As shown in Fig.~\ref{fig-DA},  the number of transmissions is effectively reduced in the hierarchical multi-layered architecture, so the energy-efficiency of the network is improved.  Data aggregation techniques depend great deal on the network architecture~\cite{survey-DA}; nevertheless,  we use a simple and common model for data aggregation: each intermediate node (e.g., a CH) aggregates all the received packets into a single output packet.  To do so, the packets should wait for a while at intermediate nodes in order to the data of other nodes are received. The operational time in EEMA is segmented into some rounds and at the beginning of each of which clusters are formed and other network operations are performed in the remaining.

The model for energy dissipation is derived from the first radio model proposed in~\cite{Heinzelman2002}.  
Accordingly, the energy needed to transmit a $l$-bit packet to distance $d$ is,
\begin{equation}
\label{eq:send-energy}
E_{t}=
\begin{cases}
l(E_{el} + \epsilon_{fs}d^{2}) \;\;\;\;\;\; d \leq d_{0}, \cr
l(E_{el} + \epsilon_{mp}d^{4}) \;\;\;\;\;\; d > d_{0},
\end{cases}
\end{equation}
where $E_{el}$ is the electronics energy, $\epsilon_{fs}$  and $\epsilon_{mp}$ are the amplifier energy of free space and multi-path models, respectively.  Also, to receive a $l$-bit packet a node consumes
\begin{equation}
\label{eq:receive-energy}
E_{r}=lE_{el}.
\end{equation}

\section{Proposed Energy-Efficient Multi-layered Architecture (EEMA)}
\label{sec:EEMA}
\begin{figure}[h]
\centering
\includegraphics[width=\linewidth]{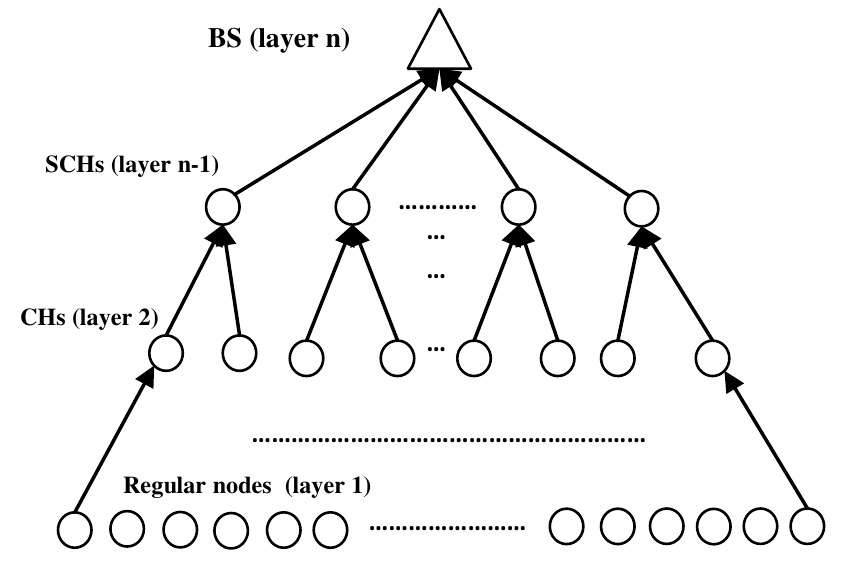}
\caption{Hierarchical architecture in EEMA. The BS is located at layer $n$ (root) and the regular nodes are located at layer 1 (leafs).}
\label{fig:CHA}
\end{figure}

In this section, we describe our proposed EEMA protocol in detail. As mentioned earlier, EEMA uses an adaptive method to form the clusters.  Term `adaptive' here means that the number of clusters varies in each round and head re-election is performed at the beginning of the rounds.
Unlike other clustering approaches in which first the clusters are formed and then the multi-hop paths are established, EEMA uses a method inspired from connection oriented services in the network layer of OSI model, wherein first all the layers are established and then the data are transmitted through a predetermined multi-hop path to the BS.  Indeed, EEMA composes an aggregation tree in which the BS is located at the {\it root} and the regular nodes are {\it leafs}, depicted in Fig.~\ref{fig:CHA}.   In the followings, first the cluster and super cluster formation algorithms are explained. Afterwards, we clarify how the data are transmitted to the BS.

\subsection{CH Election Algorithm}
\label{subsec:CH-el}

 In this section, we describe the way EEMA selects the CHs among the regular nodes.  The main idea is to select the nodes with a high residual energy, centrality, and a smaller distance to the BS as the CHs.  To do so, we introduce a novel probability for selecting the CHs as 
\begin{equation}
\label{eq-PCH}
P_{CH}(i)=\frac{E_{res}}{E_{max}}\frac{\sum_{j=1}^{k}(d_{(i,j)})}{k}\frac{d_{max}}{d_{(i,BS)}},
\end{equation}
where $E_{res}$ and $E_{max}$ indicate the residual energy and the initial energy of node $i$, respectively, $k$ is the number of neighbors within cluster range ($R_{c}$) of node $i$, $d_{(i,j)}$ indicate to the \textit{Euclidean} distance between nodes $i$ and $j$, and $d_{max}$ and $d_{(i,BS)}$ are the distance of the farthest node in the network and node $i$ to the BS, respectively. In fact, this probability assures those nodes with a higher residual energy, are more closer to the center of a dense population of the nodes, as well as to the BS have higher probability than other nodes so that which are elected as new CHs. 
At the beginning of the operations, all the nodes calculate their $P_{CH}$  and then broadcast a {\it CH-Inf} message to all neighboring nodes within $R_{c}$ to which $P_{CH}$ and the node ID are included.  Afterwards, each node waits for $t_{w}$ seconds to receive this message from all its neighbors.  Having received this message, the node first checks it and then compares $P_{CH}$ of the message with its $P_{CH}$. If the node find $P_{CH}$ of the received message greater than its $P_{CH}$, which must wait to receive the {\it CH-ADV} message from its neighbors.  Otherwise, i.e. the node waits and its $P_{CH}$ is greater than that of all the received messages, it elects itself as new CH  and  broadcasts a {\it CH-ADV} within its $R_{c}$. Note that each head node attaches its layer number to its advertisement. Note also that if a node receives no {\it CH-Inf} from its neighbors, it elects itself as new CH and broadcasts the {\it CH-ADV} within its $R_{c}$. Finally, the duration of $t_{w}$ should be reasonable; not very long to increases the time overhead of clustering, and not very short so that the nodes can receive {\it CH-Inf} from all of their neighbors.
 
 \subsection{SCH Election Algorithm}
\label{subsec:SCH-el}

After CH selection, some nodes should be selected as the super cluster-head (SCHs), as follow. Since the SCHs should aggregate the received data from bottom layers, such data are very important. Hence, the SCHs should have a proper energy level to prevent missing the data. Accordingly, we introduce a novel weight for SCH selection, as following:   

\begin{equation}
W_{SCH}(i)=(\frac{E_{res}}{E_{max}})(d_{H}(i)),
\label{eq:SCH}
\end{equation}
where $d_{H}$ indicates the number of bottom-layer head nodes which are located in the neighborhood of node $i$ that computes it by receiving the related head advertisements.  This weight assures that the nodes with a high residual energy and more proximity to bottom-layer head nodes will be selected as new SCHs. As mentioned before, the proximity to bottom-layer head nodes is important because data aggregation is effective when the data are the same, approximately. 
 Afterwards, all the nodes (except selected head nodes) compute this weight and accordingly set a timer with the value of
\begin{equation}
\label{eq-SCHT}
T_{SCH}(i)=\frac{\alpha}{W_{SCH}(i)},
\end{equation}
where $\alpha$ is a constant. Once a node's timer has expired and the node receives no {\it SCH-ADV} from its neighbors,  this node will elect itself as new SCH so that which broadcasts the {\it SCH-ADV} within $R_{s}$ (super cluster range). Note that broadcasting the {\it SCH-ADV} is performed in higher power levels, i.e. $R_{s}$ should be enough greater than $R_{c}$ in order to make sure all bottom-layer head nodes are covered.

These operations are recursively repeated until reach the BS. That is, when all the nodes in the network know their parents as well as the SCHs of the highest layer reach the BS, the SCH selection process is halted and data transmission is then began.  This is worthwhile to be mentioned, for the sake of collision avoidance, the communication between all the nodes in this phase is performed by CSMA/CA MAC layer protocol.  Afterwards, each node should join the closest head node of above-layer to itself (i.e. should join its parent) via sending a {\it Join-Req} message.  Each node finds its parent by RSSI. Head nodes after receiving these messages, add the node specifications to their members table. Note that, for the sake of fault-tolerance, each node takes a backup head node. This is because if a head gets faulty whose path could be recovered by other paths successfully. Distributed pseudo code of the proposed EEMA is presented in Algorithm~\ref{EEMA}.

\begin{algorithm}
\caption{Distributed pseudo code of head election at node $i$ by EEMA}
\begin{tabbing}
\label{EEMA}
\\
 CH El\=ection~Algorithm \\
\\
1. \> calculate $P_{CH}(i)$\\
2. \> broadcast the CH-Inf within $R_{c}$\\
3. \> wait for $t_{w}$ seconds to receive the CH-Inf\\
4. \> IF t\=he CH-Inf message is received THEN\\
5. \>\> evaluate the received messages\\
6. \>\> IF \=$\forall j, P_{CH}(i) \geq P_{CH}(j)$ THEN\\
7. \>\>\> broadcast the CH-ADV within $R_{c}$\\
8. \>\> ELSE\=\\
9. \>\>\> wait $t_{w}$ seconds to receive the CH-ADV\\
10. \>\> ENDIF\\
11. \> ELSE\=\\
12. \>\> broadcast the CH-ADV within $R_{c}$\\
13. \> ENDIF \\
\\
 SCH E\=lection~Algorithm \\
\\
1. \>IF t\=he current node is non-head THEN\\
2. \>\> calculate $W_{SCH}(i)$ and wait for $T_{SCH}(i)$ seconds\\
3. \>\> IF t\=he SCH-ADV is received THEN\\
4. \>\>\> give up the competition\\
5. \>\> ELSE\=\\
6. \>\>\> broadcast the SCH-ADV within $R_{s}$\\
7. \>\> ENDIF\\
8. \> ELSE\=\\
9.\>\> wait to receive the SCH-ADV\\
10.\>\> IF the SCH-ADV is received THEN\\
11.\>\>\> send the Join-Req to the closest above-layer SCH\\
12.\>\> ENDIF\\
13. \> ENDIF\\

\end{tabbing}
\end{algorithm}

\subsection{Data Transmission}
\label{subsec:cl-fo}

After cluster and super cluster formation, each node senses the environment and then sends its data to associated CH located in the second layer, as well as each CH gathers the data from regular nodes and aggregates and send them to its SCH in the third layer, and finally, the SCH gathers and aggregates the received data with its own data and sends to the SCH in the fourth layer, and so forth. Note that data gathering could be performed using either a TDMA protocol, with long sleep time for the regular nodes, or an on-demand approach and by the BS requests from a particular region. In general, this depends on the application and EEMA can handle both of them properly.

\section{Theoretical Analysis}
\label{sec:theo-anal}

In this section, some metrics introduced in this paper are discussed.  First of all, we opted to investigate a problem in order to simply formulate the delay between source (any node in the network) and the destination (the BS).  We know the delay between the nodes in the wireless multi-hop communications depends great deal on the MAC layer specifications~\cite{hanzo-delay}; however,  for the sake of simplicity in our evaluation, we consider  the delay as the time a packet takes to be transmitted between a source and the destination. Note that in a single-hop communication, this time is only related to the distance between the pair of source and destination; nonetheless, we are using a multi-hop method so that this time is composed of two parameters: (1) {\it link time ($t_{l}$)}, the time a packet takes to be transmitted between a pair of nodes; (2) {\it process time ($t_{p}$)}, the time a packet takes to be  processed by a node.  Note that $t_p$ includes the buffering time as well.  If $m$ indicates the number of nodes which participate in the routing, then there are ($m-1$) links among a source and the destination.  Thus, the link delay of a multi-hop path between node $i$ and the destination is calculated as
\begin{equation}
\label{eq:link-delay}
D_{l}(i)=\sum_{k=1}^{m-1}t_{l}(i,j),
\end{equation}
where $j$ is a next-hop node in the path.  As mentioned before, each node communicates with its parent so that the distance between them does not traverse $R_{s}$.  Therefore for the sake of simplicity,  if we consider $\forall j, d_{(i,j)} \leq R_{s}$, then equation~(\ref{eq:link-delay}) could be simplified as
\begin{equation}
\label{eq:flink-delay}
D_{l}(i)=(m-1)t_{l}.
\end{equation}
Similarly, there are ($m-2$) intermediate nodes in the path which should process the packet to route it. Thus,
\begin{equation}
\label{eq:process-delay}
D_{p}(i)=\sum_{k=1}^{m-2}t_{p}(j).
\end{equation}
Since used network is homogeneous and all the nodes have the same capabilities, including the processing power, so equation~(\ref{eq:process-delay}) simplifies to
\begin{equation}
\label{eq:fprocess-delay}
D_{p}(i)=(m-2)t_{p}.
\end{equation}
The total delay between nodes $i$ and $j$ in the network can be achieved by combining equations~(\ref{eq:link-delay}) and (\ref{eq:process-delay})
\begin{equation}
\label{eq:total-delay}
D_{t}(i,j)= \sum_{k=1}^{m-1}t_{l} + \sum_{k=1}^{m-2}t_{p}.
\end{equation}
In our scheme, this delay may be achieved by
\begin{equation}
\label{eq:tot-delay}
D_{t}(i,BS)=(m-1)t_{l} + (m-2)t_{p}.
\end{equation}
Although EEMA imposes a trivial overhead for SCH election to the clustering algorithm, which removes the routing delay.  According to equation~\eqref{eq:total-delay}, routing in large-scale WSNs with a huge number of nodes incurs a significant delay, specially in flat architectures.  This delay is not acceptable in many applications of WSNs in which a quick response is required.  We further discuss it in section~\ref{sec:exp}. 

As discussed earlier,  some nodes in the architecture of EEMA are elected as the SCHs.  These nodes are very important because they possess all the data of the network.  Since the number of head nodes in EEMA is more than that in clustering protocols, a problem is that whether the head nodes should participate in sensing the environment or not.  Clustering approaches usually try to cluster the network with as few cluster as possible so that the number of the CHs compared to the number of all the nodes in the network is approximately negligible.  Thus, the CHs, which perform many energy-consumer tasks in the network (receiving, aggregating and transmitting the data), are not required to sense the environment.  On the other hand, since each SCH is responsible to receive and aggregate the data of a few CHs of its super-cluster, EEMA enforce them to participate in sensing the environment so that the coverage is not affected by our approach.  I.e, EEMA conserves the coverage of the field as good as clustering approaches.  More precisely,  we study this by formulating energy consumption of the CHs and then compare it with that of the SCHs.  Consider a CH has $n_{c}$ cluster members within its cluster.  The CH has the following energy consumption rate  to perform its tasks:
\begin{equation}
\label{eq:E_ch}
E_{ch}=E_{r}(n_{c}-1) + lE_{da}n_{c} + E_{t}, 
\end{equation} 
where $E_{da}$ is the energy for data aggregation.  Similarly, a SCH with $n_s$ super-cluster members has to consume the following energy to perform its tasks:
\begin{equation}
\label{eq:E_sch}
E_{sch}=E_{r}(n_{s}-1) + lE_{da}n_{s} + E_{t}. 
\end{equation} 
Here, the number of members within each cluster and super-cluster determines energy consumption in the CHs and SCHs.  Since the nodes are distributed randomly and uniformly, so
\begin{equation}
\label{eq:density}
N=\lambda \times |A|=\lambda M^2. 
\end{equation}
According to equation~\eqref{eq:density}, $n_{c}$ is achieved by
\begin{equation}
\label{eq:n-c}
n_c=\lambda\pi R_{c}^2. 
\end{equation}
On the other hand, in order to compute $n_s$ we need to know the number of clusters in the network or the density of CHs. 
If $k_c$ is the number of clusters in the network, we have 
\begin{equation}
\label{eq:cluster-num}
k_c=\frac{M^2}{\pi R_{c}^2}. 
\end{equation}
Also, this number for super-clusters is
\begin{equation}
\label{eq:sup-cluster-num}
k_s=\frac{M^2}{\pi R_{s}^2}. 
\end{equation}
Now, $n_s$ is computed by dividing equations~\eqref{eq:cluster-num} and~\eqref{eq:sup-cluster-num}
\begin{equation}
\label{eq:n-s}
n_s=\frac{k_c}{k_s}=\frac{R_{s}^2}{R_{c}^2}. 
\end{equation}
For example, if we take $N=4000$ and $M=1000$, then $\lambda=0.004$.  Assuming $R_c=50m$ and according to equation~\eqref{eq:n-c}, $n_c=31$.  As discussed earlier, $R_s$ should be greater than $R_c$.  On the other hand, in order to conserve the connectivity $R_s$ must be less than $R_t$.  Consider $R_s=100m$, so $n_s=4$, according to equation~\eqref{eq:n-s}.  Therefore, according to equations~\eqref{eq:E_ch} and~\eqref{eq:E_sch}, each CH has to consume about 8 times more energy than each SCH and our previous saying, i.e. SCHs have to participate in data acquisition, seems logical.

In order to accurately define $R_{s}$, we should consider the connectivity condition.  As discussed in~\cite{PEAS2003}, in order to a multi-hop network remains connected, probing range should be sufficiently less than the maximum transmission range of the node.  If we take $R_c$ equals the probing range in~\cite{PEAS2003}, the sufficient condition for connectivity is $R_{t}>6R_c$ (for proof see~\cite{PEAS2003}).  Accordingly, we can define a limit for $R_s$, as the below
\begin{equation}
\label{eq:limit}
R_c \leq R_{s} < 6R_{c}. 
\end{equation}
Defining optimal $R_s$ depends on finding the optimal number of SCH across the network.  Similar to finding the optimal number of clusters in the network, finding the optimal number of the SCH in the network is a NP-hard problem~\cite{brief-sur}.  As discussed earlier, some works like~\cite{karaki2004}\cite{karaki2009} try to solve this problem.  However, since multi-layered architectures are usually employed in the large-scales of WSNs, and also finding the optimal solution using heuristic algorithms needs a global information of the network (the centralized approach), so this method has practically problem.  In distributed systems, mathematical optimization techniques have gained more popularity.  In EEMA, the optimum number of head nodes depends great deal on the optimum number of layers in the network.  It is obvious that the number of layers grows up with increasing the network scales.  On the other hand, the number of layers depends on the maximum transmission range of a node.  More precisely, each layer has own itself $R_s$.  For example, if $R_c=50m$, $R_s$ in 3rd layer equals $100m$, 4th layer $150m$, etc.  In the last layer, $R_s=R_t$ in order to find more bottom-layer head nodes.  When $R_s$ reached $R_t$, no more layer would be possible.  Once the number of layers in the network gets apparent, the optimal number of head nodes in each layer could be found by one of~\cite{Heinzelman2002}\cite{opt-num-clu2004}\cite{optimalwang2009} that pursue finding the optimum number of head nodes in a two-tier architecture.  We further analyze this by simulation in the next section.

The incurred overhead, in terms of message and time, by EEMA is acceptable compared to clustering approaches.  Firstly, we study the message and time complexity of clustering by EEMA.  As explained earlier, each CH node has to send two messages at most, i.e. {\it CH-Inf} and {\it CH-ADV}, as well as each regular node two messages, i.e. {\it CH-Inf} and {\it Join-Req}.  Thus, considering there are $N$ nodes in the network, all the transferred messages for CH election is $4 \times N$, consequently, the message complexity would be the order of $O(n)$.  Also, each node in the network should wait for $t_w$ seconds, and since this waiting is performed in parallel, so the time complexity is also the order of $O(n)$.  Secondly, for SCH election, EEMA conserves the message and time complexity the order of $O(n)$, as discussed in the following.  The number of additional messages produced by EEMA depends on the number of the SCHs.  As discussed previously, this number is surely less than that of the CHs.  Each SCH broadcasts a {\it SCH-ADV}, and consequently, whose members send to which a {\it Join-Req} message.  Thus, assuming the number of SCHs and CHs equals $k_s$ and $k_c$, respectively, the number of transferred messages for SCH election is
 \begin{equation}
\label{eq:mes-comp}
n_m=k_s + [(k_s + k_c) - k_l]= 2k_s + k_c - k_l,
\end{equation}
where ($k_s + k_c$) indicates the number of head nodes should send {\it Join-Req} message, and $k_l$ indicate to the head nodes of the last layer which are not required to send {\it Join-Req}.  This conveys that the message complexity of EEMA is the order of $O(n)$.  Also, this should be mentioned that EEMA reduces the whole load of the network by effective aggregating.  As the time complexity, the only extra time that EEMA imposes to the network is the waiting time for SCH election.  In the worst case, this time is the order of $mS$ that comparing to the entire operation time of the network is negligible.  

As discussed earlier, the synchronization could locally achieved.  The most important part of the algorithm that urgently needs the nodes to be synchronized is the first of each round.  Full synchronization in large distributed systems is very tough, even practically impossible.  However, since our architecture is hierarchical cluster-based, local synchronization might be achieved through a few bytes packet exchange~\cite{zhu}.  This message exchange is worth because which avoids message collision in the network.

Head election in EEMA is basically energy-aware.  This is very important because the head node possess all the data of the network, and the dying of them means losing these data.  The introduced weight for SCH election has a parameter shown by $d_{H}$.  As mentioned earlier, data aggregation is more effective when the data is aggregated from the same regions, approximately.  To do so, $d_{H}$ is wisely employed in the weight to select the nodes with a higher degree of bottom-layer head nodes as new SCHs.  On the other hand, timer-based SCH election results in reducing the required message for SCH election so that the network is more energy-efficient.

Finally, EEMA handles both types of data acquisition methods in WSNs, i.e. TDMA-based or on-demand.  TDMA-based WSNs are usually employed in the applications of monitoring a static environment.  A good illustration of that can be monitoring a farm in agricultural applications~\cite{survey-app} or habitat monitoring~\cite{habitat}.  On-demand approach is operated in applications in which the BS has many interactions with the nodes.  Military applications of WSNs could be one of such applications~\cite{survey-app}.  In general, EEMA properly manages both of them.  This is worthwhile to be mentioned that in applications that a more reliable approach is needed, EEMA can enforce the nodes to take more than one parent.  This causes the network remains fully-connected, and potentially, which is fault-tolerant~\cite{paradis2007}.  We leave this for our future work.

\section{Experiments}
\label{sec:exp}

In this section, we first describe the simulation setup, and then, the results of simulations are given.

\subsection{Simulation Setup}
\label{subsec:sim-set} 

We use a large WSN with the following scenarios:
\begin{itemize}
\item {\bf First scenario:}  300 nodes that are uniformly and randomly dispersed in a field of size $1000m \times 1000m$;

\item  {\bf Second scenario:} 1000 nodes that are uniformly and randomly dispersed in a field of size $2000m \times 2000m$.

\end{itemize}
We assume that the BS is located at the center of the field.  We have considered the network lifetime as three metrics~\cite{handy2002}:
\begin{itemize}
 
\item {\bf FND} (First Node Dies): Interval between the start of the operations until the first node dies.
\item {\bf HNA} (Half of the Nodes Alive): Interval between the start of the operations until half node dies.
\item {\bf LND} (Last Node Dies): Interval between the start of the operations until the last node dies.

\end{itemize}
 All the results are the average of over 50 runs.  The energy model is taken from~\cite{Heinzelman2002}.  For the sake of comparison, EEMA is compared with some well-known and state-of-the-art clustering protocols: LEACH, HEED, DWEHC, and EEDC.  We have modified LEACH to support multi-hop communications through data transmission to the BS.  Also, the simulations results of LEACH is the average of results when the CH probability ($p$) in which varies from $0.05$ to $0.15$.  We have simulated HEED with AMRP factor.  For DWEHC, we let the approach supports multi-hop intra-cluster communications.  Finally for EEDC, the threshold distance $D_{thr}=30$ and the local competition range $R_{comp}=25$.
 Particularly, we have selected HEED and DWEHC for comparison, because they have used a large-scale WSN in their simulations.  For the sake of simplicity and reality in delay evaluation, since the exact evaluation depends on MAC layer specifications of the nodes, we have used the following equivalents.  We have considered the transmission over one meter as one time unit ($t_u$); also, we have taken the time between the arrival of a message into a node until which leaves the node equals $10 \times t_u$.  Other simulation parameters are summarized in Table~I.

\begin{table}[h]
\centering
\caption{Simulation Parameters}
\label{tab:1}       
\begin{tabular}{l l l}
\hline\noalign{\smallskip}
Parameter & Scenario 1 & Scenario 2 \\
\noalign{\smallskip}\hline\noalign{\smallskip}
BS location & $(500,500)$ & $(1000,1000)$\\
$M$ & $1000m$ & $2000m$\\
$N$ & 300 & 1000\\
$\epsilon_{fs}$ & $10pJ/bit/m^2$ & $10pJ/bit/m^2$\\
$\epsilon_{mp}$ & $0.0013pJ/bit/m^4$ & $0.0013pJ/bit/m^4$\\
$E_{el}$ & $50nJ/bit$ & $50nJ/bit$\\
$E_{da}$ & $5nJ/bit/signal$ & $5nJ/bit/signal$\\
Initial Energy & $8J$ & $8J$\\
Data Frame & $500Byte$ & $500Byte$\\
$d_{0}$ & $87m$ & $87m$\\
$R_{c}$ & $50m$ & $50m$\\
$R_{s}$ & $100ms$ & $100ms$\\
$R_{t}$ & $300ms$ & $300ms$\\
\noalign{\smallskip}\hline
\end{tabular}
\end{table}

\subsection{Simulation Results}
\label{subsec:sim-res}

\begin{figure*}[t]
	\centering
	\subfigure[First Scenario.]{
\includegraphics[width=0.45\linewidth]{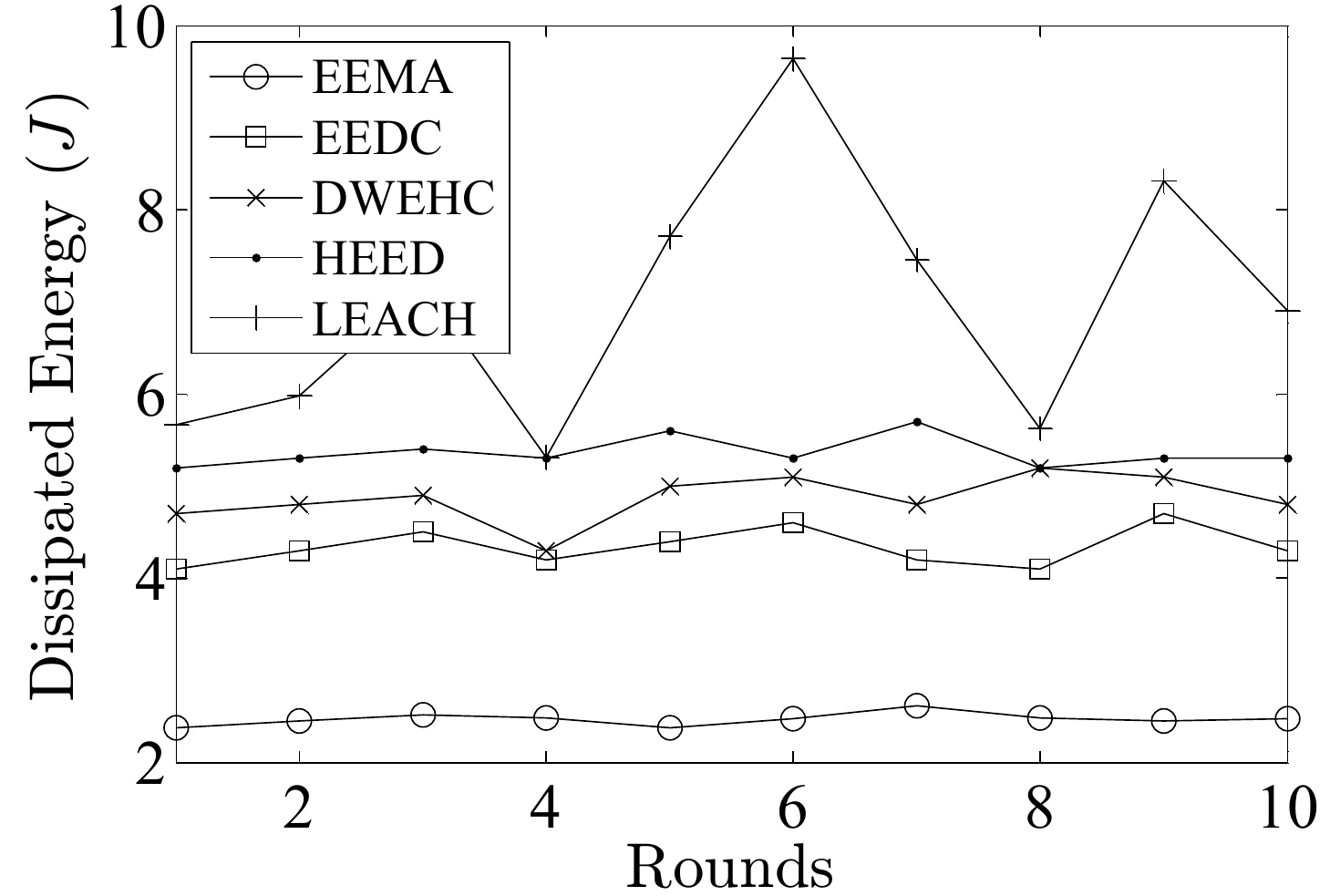}
	    \label{fig:CE-FS}
	}
    \subfigure[Second Scenario.]{
    \includegraphics[width=0.45\linewidth]{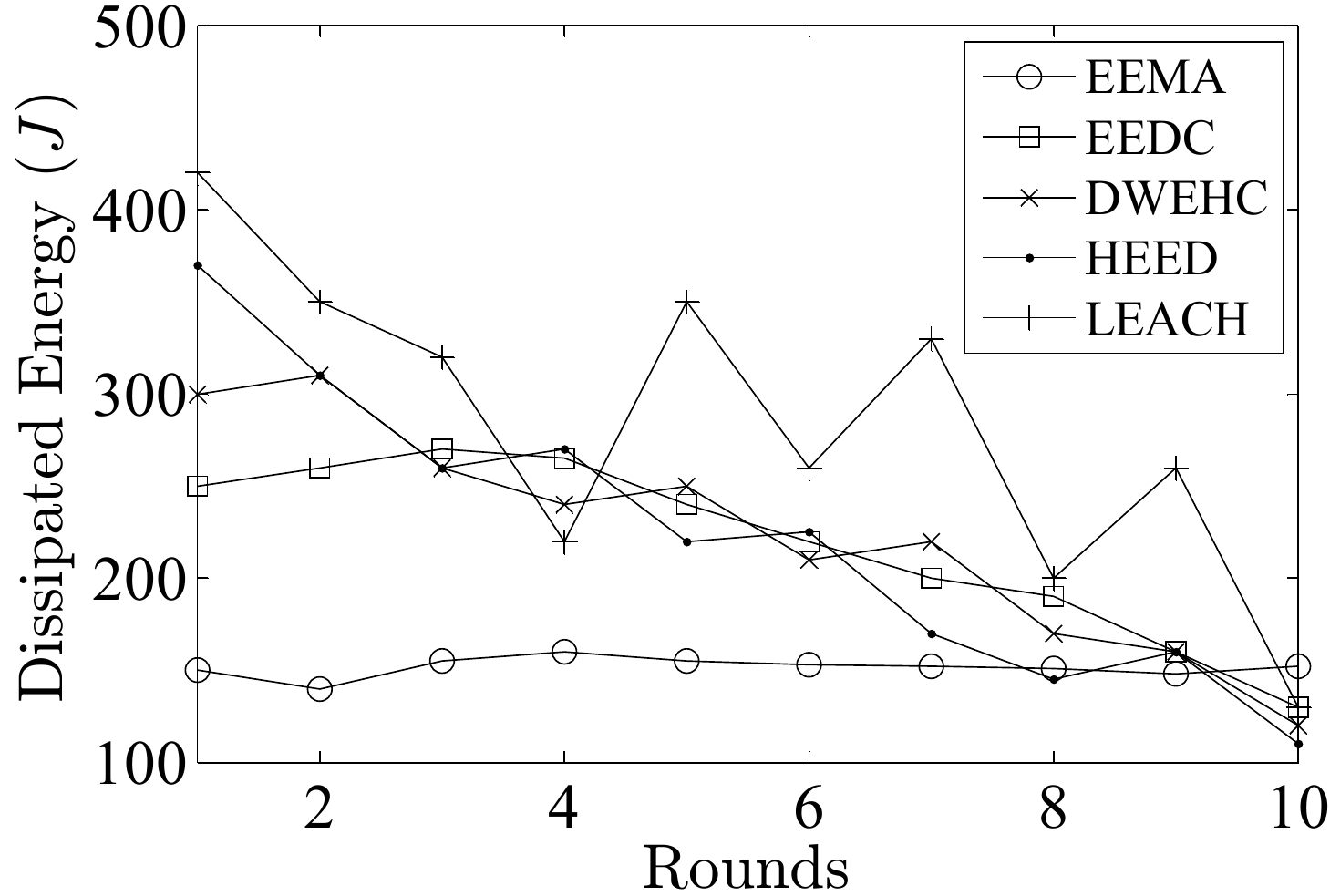}
	\label{fig:CE-SS}
	}
	\caption[Optional caption for list of figures]{The dissipated energy by all the nodes in the network.}
	\label{fig:CE}
\end{figure*}

\begin{figure*}[t]
	\centering
	\subfigure[First Scenario.]{
\includegraphics[width=0.3\linewidth]{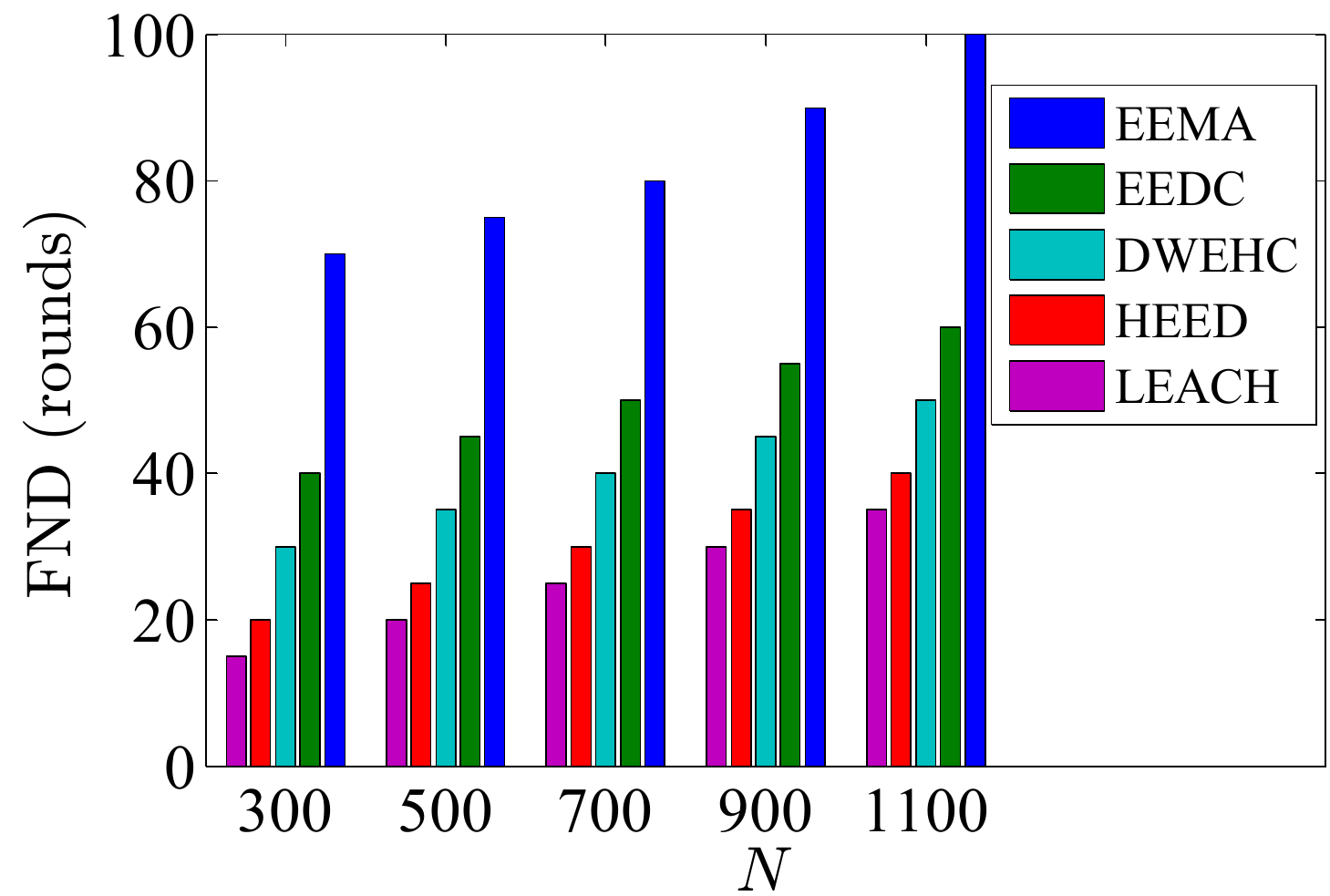}
	    \label{fig:FND-FS}
	}
    \subfigure[Second Scenario.]{
    \includegraphics[width=0.3\linewidth]{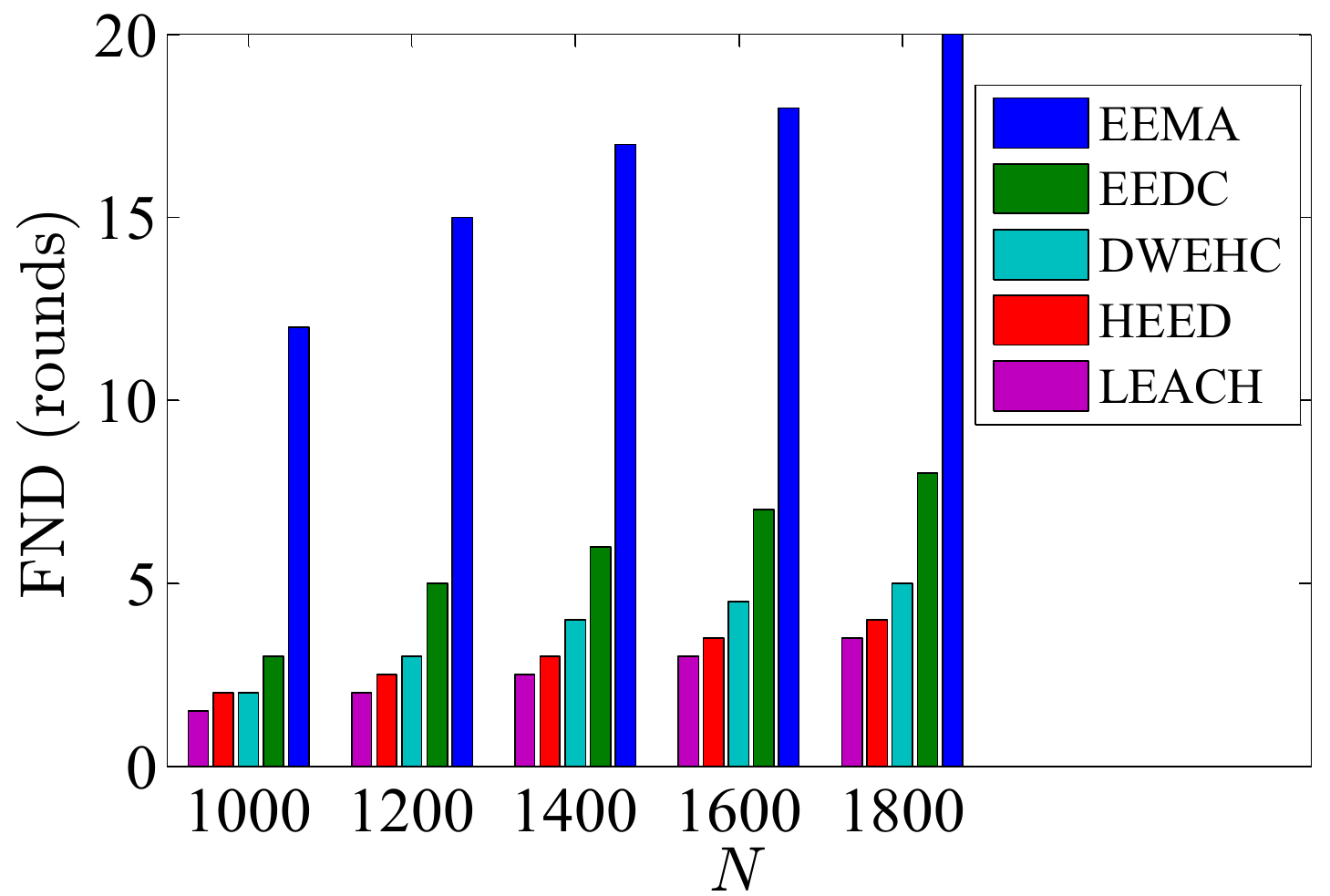}
	\label{fig:FND-SS}
	}
		\subfigure[First Scenario.]{
\includegraphics[width=0.3\linewidth]{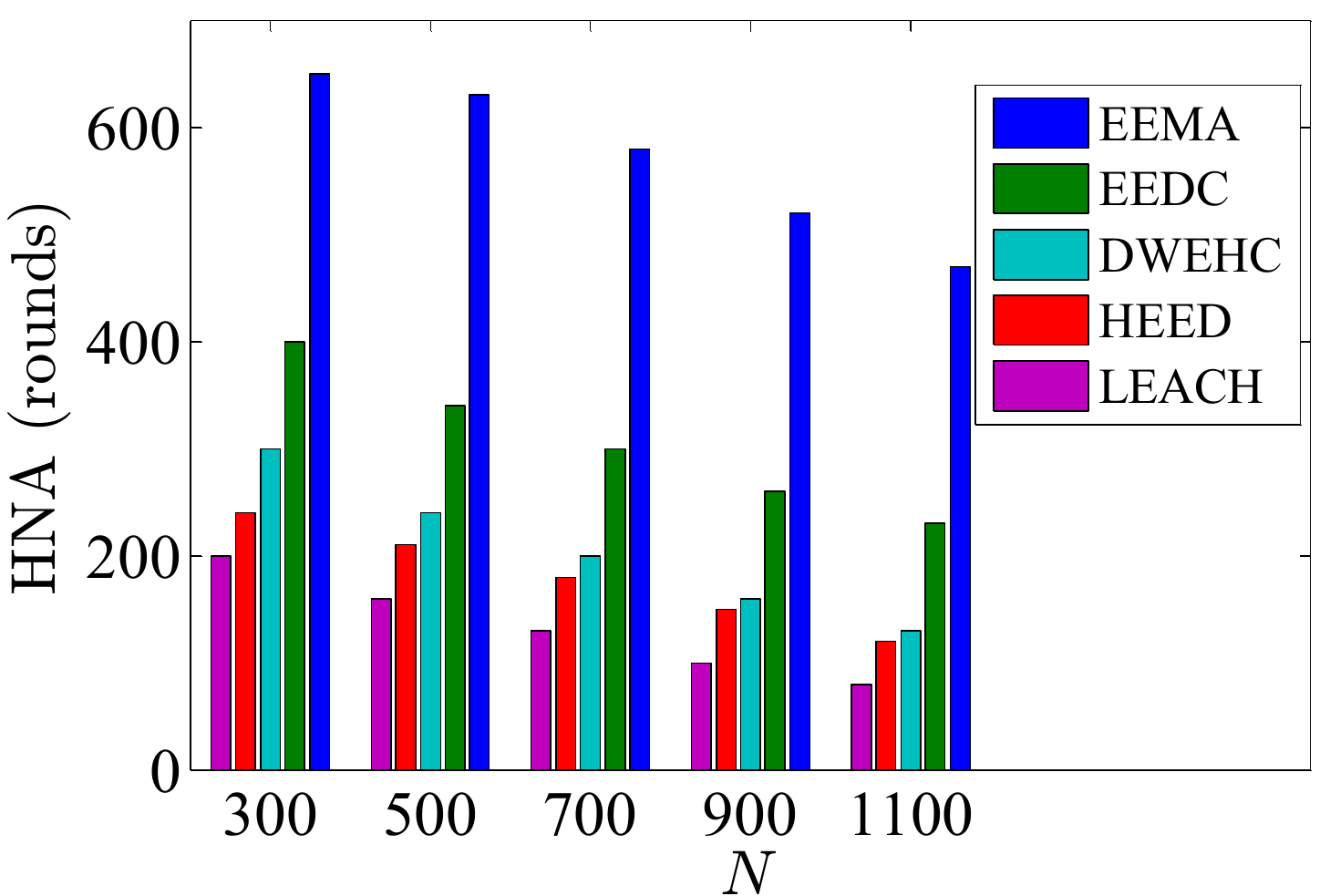}
	    \label{fig:HNA-FS}
	}
    \subfigure[Second Scenario.]{
    \includegraphics[width=0.3\linewidth]{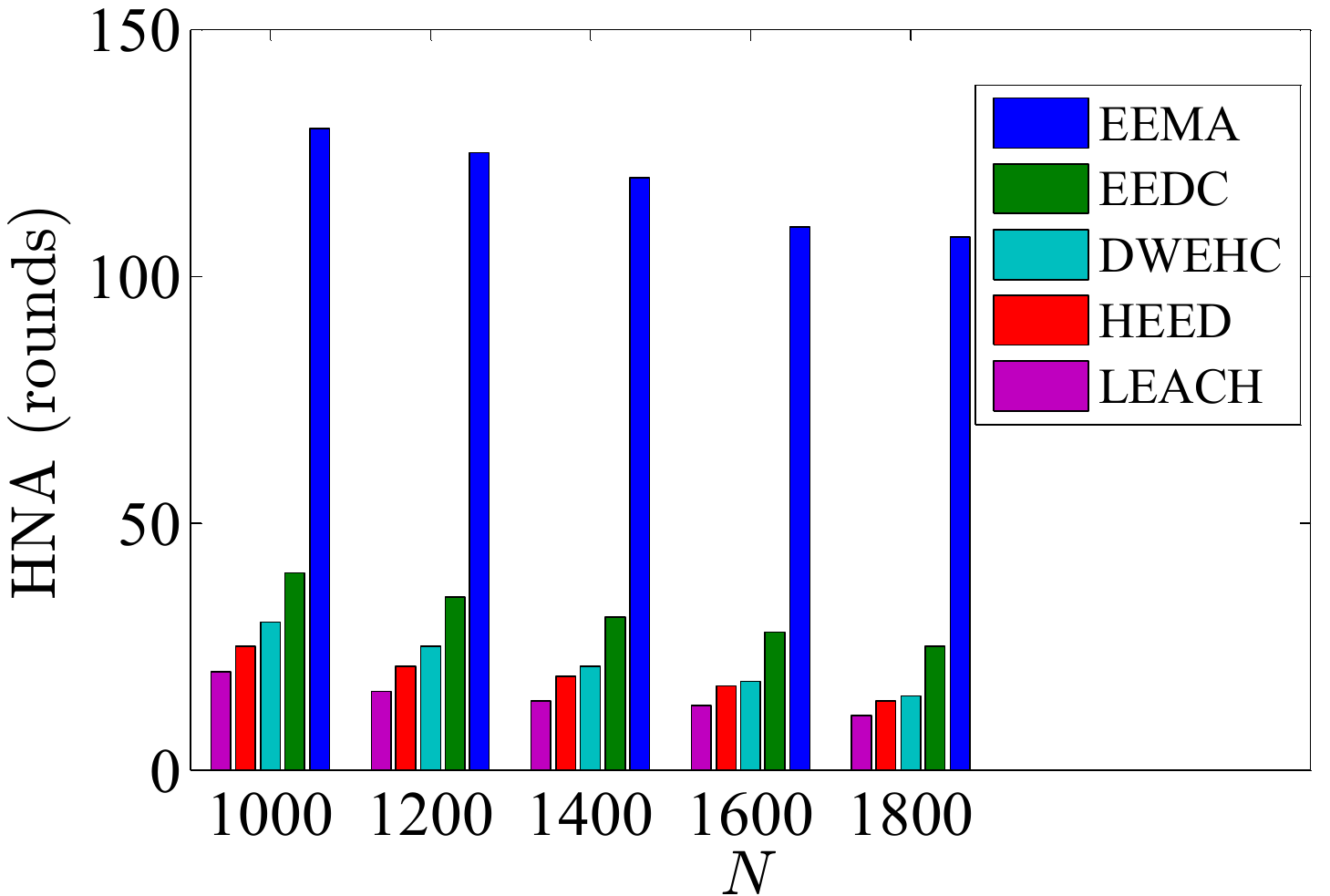}
	\label{fig:HNA-SS}
	}
		\subfigure[First Scenario.]{
\includegraphics[width=0.3\linewidth]{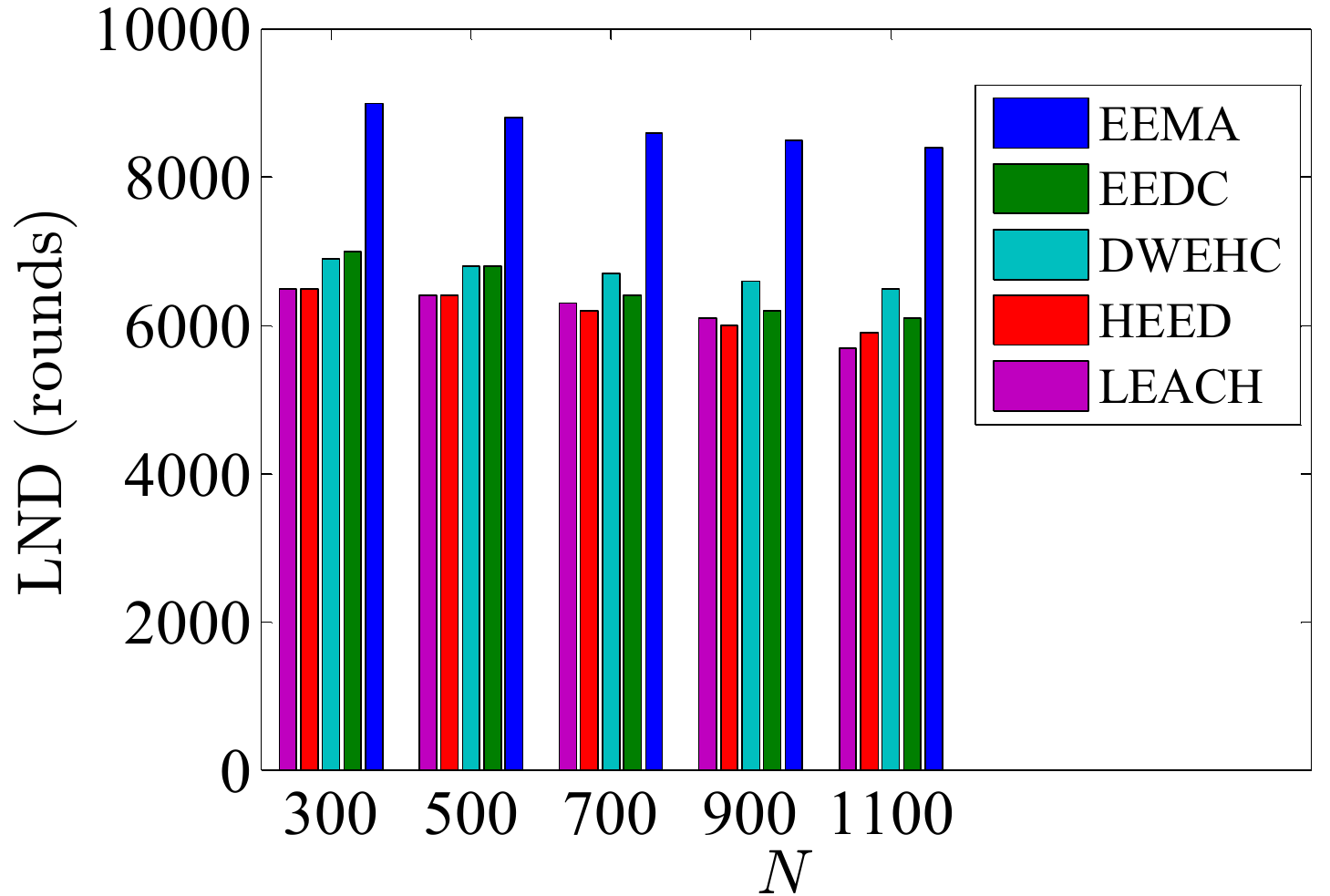}
	\label{fig:LND-FS}
	}
    \subfigure[Second Scenario.]{
    \includegraphics[width=0.3\linewidth]{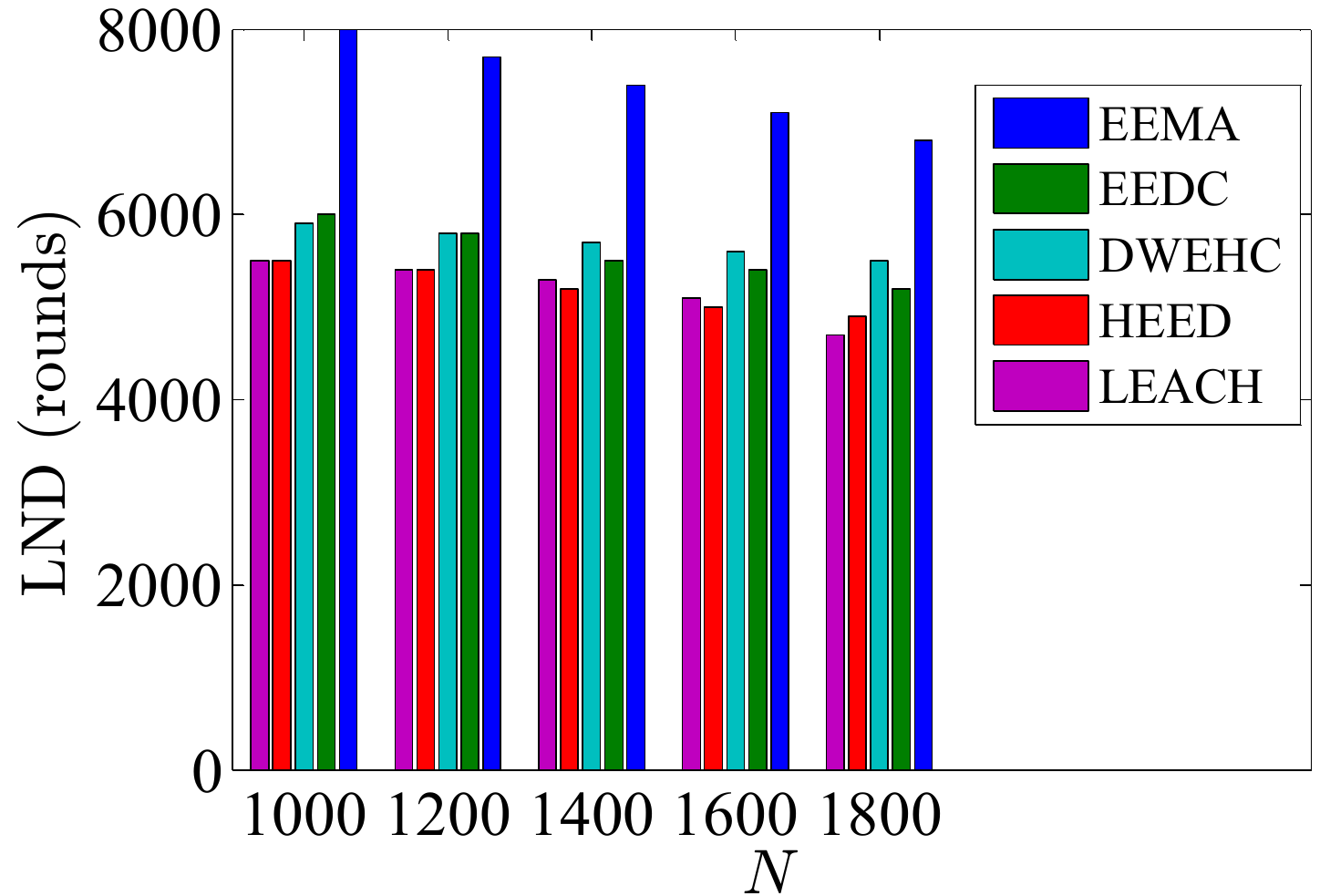}
	\label{fig:LND-SS}
	}
	\caption[Optional caption for list of figures]{The network lifetime with different metrics.}
	\label{fig:Lifetime}
\end{figure*}

In this section, the results are presented. In Fig.~\ref{fig:CE-FS} and~\ref{fig:CE-SS}, the average dissipated energy by all the nodes in the network for two scenarios is depicted.  As is seen, EEMA shows a constant and low rate of energy consumption in both scenarios.  As expected, when the scales of the network are increased, the energy consumption is increased as well.  This is because the intra- and inter-cluster communication costs are inevitably raised.  As shown, EEDC, HEED, and DWEHC have a relatively constant energy consumption rate, and also higher than EEMA.  Because of the random-based design, LEACH has a variant rate of energy consumption.  However, other protocols, i.e. EEMA, EEDC, HEED, and DWEHC, since which have an energy-aware design, this rate is relatively stable.  In the second scenario, it seems that EEDC, HEED, DWEHC, and LEACH have a better energy consumption than EEMA with the increase of the rounds, shown in Fig.~\ref{fig:CE-SS}.  For the sake of explanation, in the baseline clustering protocols, since the nodes start to die after round 3, the average consumed energy for a lower number of nodes is less than that of EEMA in which all the nodes are alive.  In general, the data suggest that EEMA consumes about 100\% less energy than the competitive clustering approaches.  The main cause of lower energy consumption of EEMA is that which effectively reduces the load of the network by selecting aggregation points (SCHs). 
The network lifetime for protocols regarding the defined metrics are presented in Fig.~\ref{fig:FND-FS} to~\ref{fig:LND-SS}.  As is observable, EEMA outperforms all the clustering protocols in terms of the network lifetime.  More precisely, regarding the FND, EEMA improves the network lifetime about by 100-150\% compared with the baseline clustering protocols.  As shown in Fig.~\ref{fig:FND-FS} and~\ref{fig:FND-SS}, EEMA shows a better performance in a more dense network.  This is because when the number of nodes is increased, the extra nodes are able to distribute the load of the network so that the network lifetime is relatively improved in contrast with a sparse network.  A higher FND is concurrent with a lower HNA (shown in Fig.~\ref{fig:HNA-FS} and~\ref{fig:HNA-SS}); this is because when the load is well distributed among all the nodes in the network, which almost have the same residual energy so that the time in which all the nodes are alive (FND) is longer.  Once the first node dies, all the the nodes which have a relatively equal energy are died consequently (almost simultaneous), and as a result, the HNA is decreased.  This is the case for the LND metric, shown in Fig.~\ref{fig:LND-FS} and~\ref{fig:LND-SS}.  

On the other hand, EEDC shows a better performance than other clustering protocols in terms of the FND and HNA.  The main reason is that EEDC opts the residual energy as the first factor and the nodes with the highest residual energy are elected as the CHs, and as a result, the FND and HNA are improved.  Although HEED and DWEHC utilize the residual energy in their CH election, the final CHs are elected based on some other metrics, including the node degree and the proximity to other nodes.  As mentioned earlier, LEACH has the lowest performance in terms of the network lifetime, because of its random nature.  In general, according to figures, although EEMA sufficiently improves the network lifetime comparing to the clustering protocols, this improvement is more eminent in the second scenario, i.e. in the larger network.  As a result, we can conclude that multi-layered architectures are more appropriate for large-scale WSNs. 

\begin{figure*}[t]
	\centering
	\subfigure[ ]{
\includegraphics[width=0.3\linewidth]{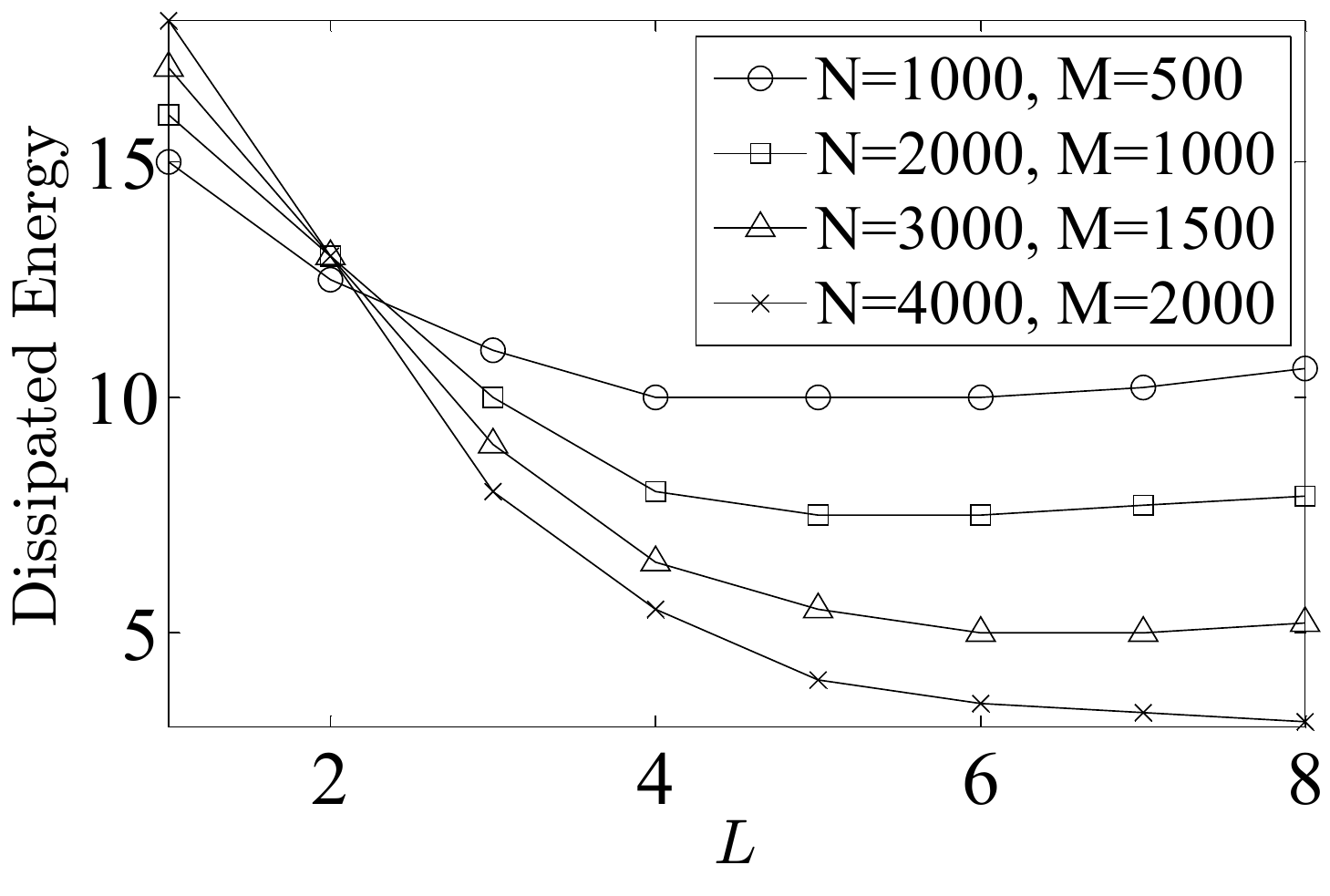}
	    \label{fig:Spent}
	}
    \subfigure[]{
    \includegraphics[width=0.3\linewidth]{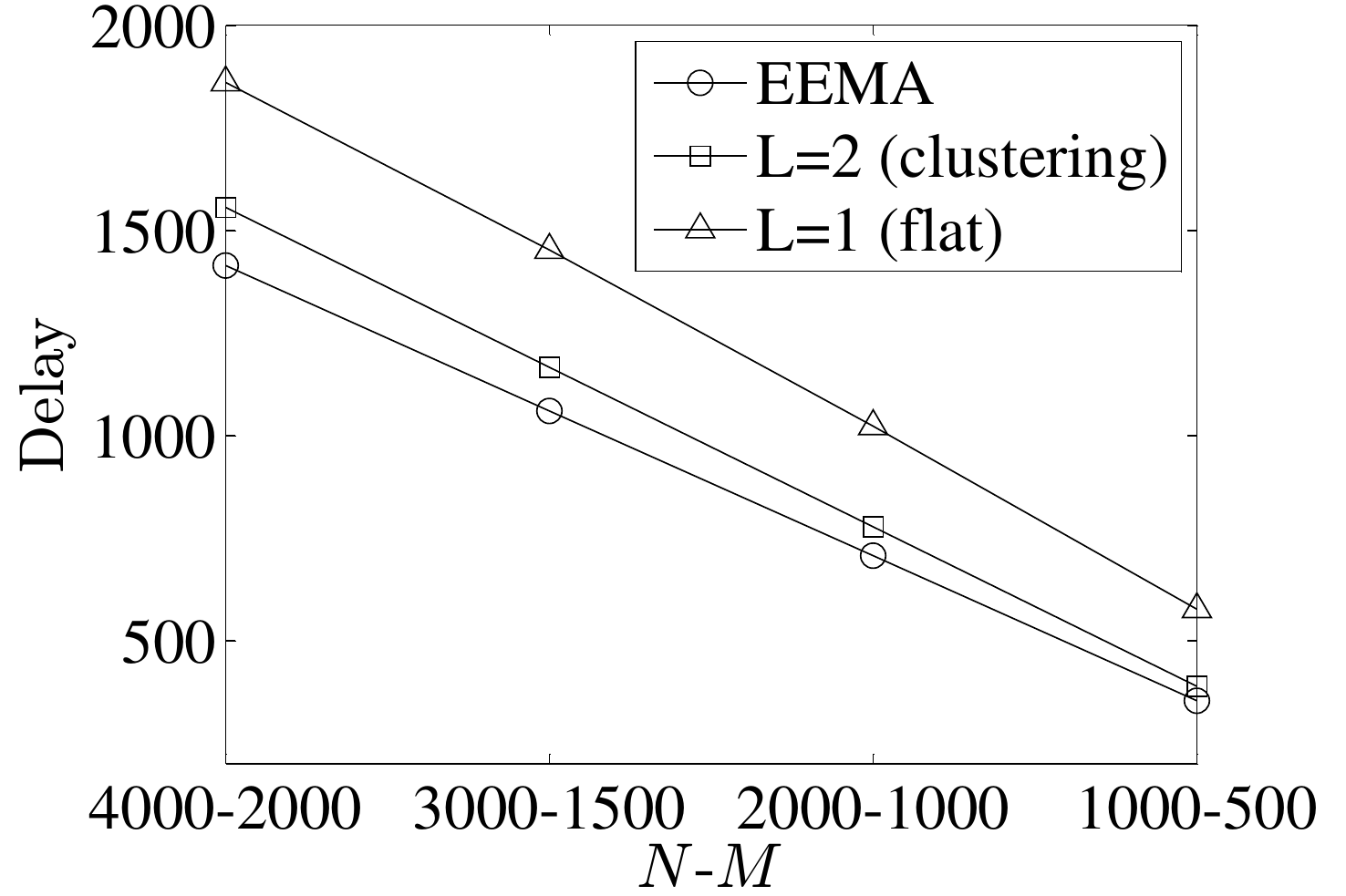}
	\label{fig:Delay}
	}
	\subfigure[]{
    \includegraphics[width=0.3\linewidth]{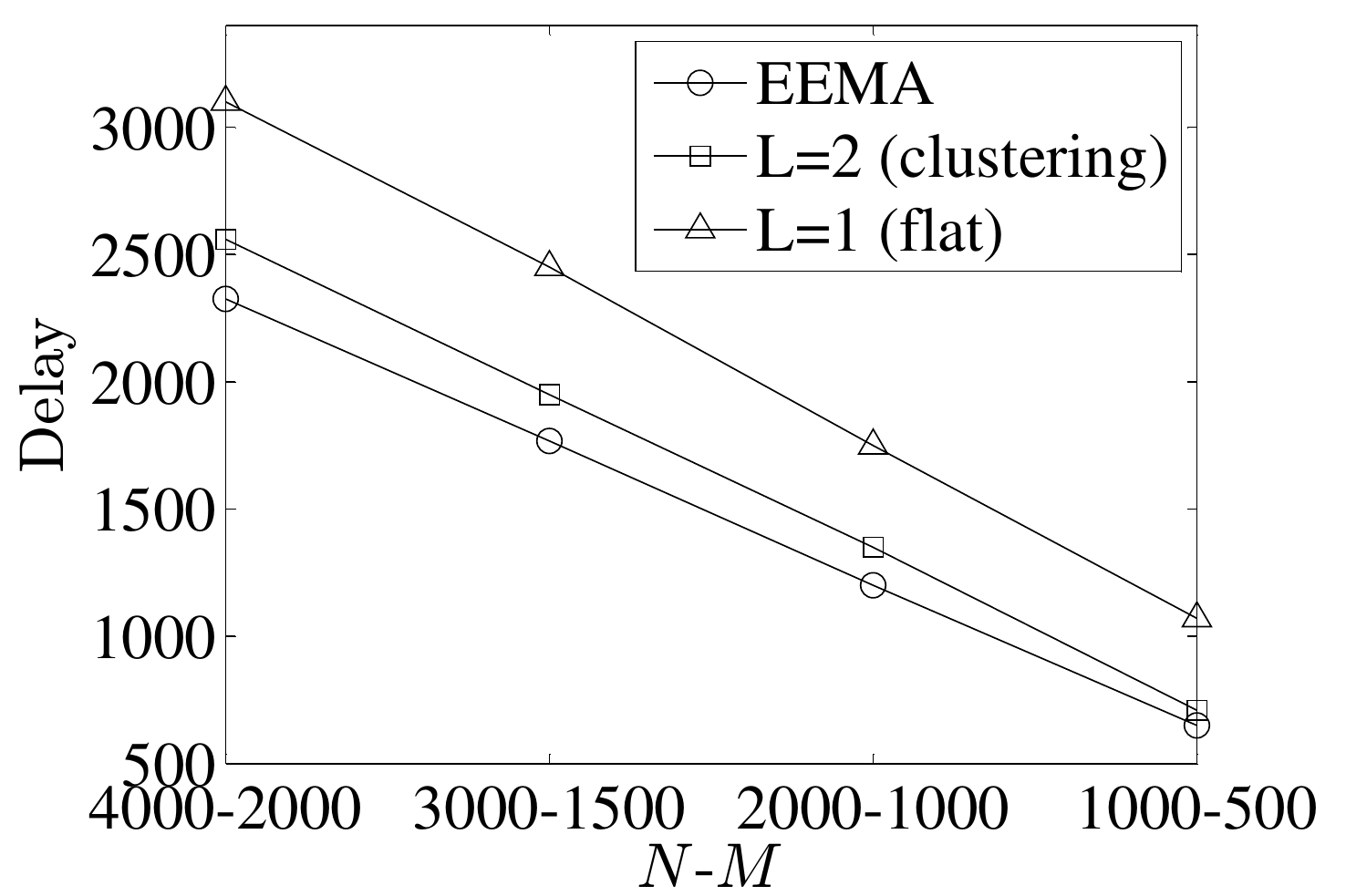}
	\label{fig:Delay-ss}
	}
	\caption[Optional caption for list of figures]{(a) Spent energy by all the nodes for adding the layers to the architecture (the spent energy is in logarithm). (b) The routing delay in different scales with the BS at the center of the field. (c) The routing delay in different scales with the BS located at ($M+100, M/2$).}
	\label{fig:ene-del}
\end{figure*}

In order to achieve a rational result for adding the extra layers to the network, Fig.~\ref{fig:Spent} is presented.  In this case, we let the network dimensions and the number of nodes to be variable between 300 to 2000 and 500 to 4000, respectively. As is seen, the spent energy in the entire network is variant proportional to different scales and layers; as when the network is smaller ($N=1000$ and $M=500m$), adding the extra layers results in a higher energy consumption rate.  Nonetheless, in the larger networks, adding the extra layers makes the network more energy-efficient.

Finally, the routing delay between source and the destination (the BS) is depicted in Fig.~\ref{fig:Delay} and~\ref{fig:Delay-ss}, where in the former the BS is located at the center of the field, and in the latter the BS is located at ($M+100,M/2$).
To evaluate the delay, we have used equation~(\ref{eq:tot-delay}) in our simulations in which the farthest node in the network to the BS is considered as the source.  As shown, the flat architecture ($L=1$) has the highest delay, because there are more nodes between the source and the BS that the message should be routed by them.  Although this delay is properly decreased in clustering architecture ($L=2$),
EEMA has the best improvement on delay because in which $t_p$ is removed (see section~\ref{sec:theo-anal}).  
According to Fig.~\ref{fig:Delay-ss}, this improvement is more considerable, as when $N=4000$ and $M=2000$, EEMA improves the delay about 30\% compared to the flat architecture, consequently,  the network would be more scalable.  Note that in hierarchical multi-layered architectures, data aggregation is performed so that the waiting delay of packets in the intermediate nodes (i.e. the CHs and SCHs) might increase the delay.  On the other hand, since data aggregation is not employed in the flat architecture, this delay is removed.  In our simulations, we have not considered the aggregation delay.

\section{Conclusion}
\label{sec:con}
This paper proposes EEMA (Energy-Efficient Multi-layered Architecture), a novel adaptive hierarchical architecture protocol for large-scale WSNs. EEMA divides the network into some virtual layers, as well as each layer to some clusters and super clusters. Basically, EEMA constructs an hierarchical aggregation tree in which the BS is located at root and the regular nodes constitute leafs.  The heads of each layer are selected proportional to their residual energy, centrality, and the distance to the BS, as well as their proximity to bottom-layer head nodes.  The results of simulations have confirmed the effectiveness of proposed EEMA for large-scale WSNs, in terms of effective data aggregation, increased network lifetime, and reduced routing delay.

\bibliographystyle{IEEEtran}
\bibliography{general1}

\end{document}